\documentclass[aip,apl,reprint,superscriptaddress,amsmath,amssymb]{revtex4-1}

\usepackage{graphicx,bm,hyperref,natbib}

\begin{document}
\title{Multi-frequency Spin Manipulation Using Rapidly Tunable Superconducting Coplanar Waveguide Microresonators}
\author{A. T. Asfaw}
 \email{asfaw@princeton.edu}
\affiliation{
    Department of Electrical Engineering, Princeton University, Princeton, New Jersey 08544, USA
}
\author{A. J. Sigillito}
\affiliation{
Department of Electrical Engineering, Princeton University, Princeton, New Jersey 08544, USA
}
\author{A. M. Tyryshkin}
\affiliation{
    Department of Electrical Engineering, Princeton University, Princeton, New Jersey 08544, USA
}
\author{T. Schenkel}
\affiliation{
    Accelerator Technology and Applied Physics Division, Lawrence Berkeley National Laboratory, Berkeley, California 94720, USA
}
\author{S. A. Lyon}
\affiliation{
    Department of Electrical Engineering, Princeton University, Princeton, New Jersey 08544, USA
}
\date{\today}
\begin{abstract} 
    In this work, we demonstrate the use of frequency-tunable superconducting NbTiN coplanar waveguide microresonators
    for multi-frequency pulsed electron spin resonance (ESR) experiments. By applying a bias current to the center pin,
    the resonance frequency ($\sim$7.6 GHz) can be continuously tuned by as much as 95 MHz in 270 ns without a change in
    the quality factor of 3000 at 2K.  We demonstrate the ESR performance of our resonators by measuring donor spin
    ensembles in silicon and show that adiabatic pulses can be used to overcome magnetic field inhomogeneities
    and microwave power limitations due to the applied bias current.  We take advantage of the rapid
    tunability of these resonators to manipulate both phosphorus and arsenic spins in a single pulse sequence,
    demonstrating pulsed double electron-electron resonance (DEER).  Our NbTiN resonator design is useful for
    multi-frequency pulsed ESR and should also have applications in experiments where spin ensembles are used as quantum
    memories.
\end{abstract} 
\maketitle 

Superconducting microresonators have dramatically enhanced the detection sensitivity of conventional electron spin
resonance (ESR). Recently, single-shot detection of $10^7$ spins has been demonstrated at 2K using coplanar waveguide
(CPW) resonators \cite{Sigillito_2014,Sigillito2017}, and further improvements in the readout at mK temperatures
have enabled detection of $10^3$ spins \cite{eichler_electron_2017,bienfait_reaching_2016} using lumped-element
resonators. As a result, these resonators are routinely used for the manipulation of spin ensembles and have also
been proposed as buses for coherently coupling superconducting qubits with spins
\cite{petersson_circuit_2012,kubo_hybrid_2011}. The resonance frequencies of CPW and lumped-element resonators are
typically fixed by fabrication, and conventional methods used to tune the resonance frequency post-fabrication using
superconducting quantum interference devices are incompatible with the high magnetic fields that are typically necessary
for X-band ESR.  A method that enables frequency tunability of superconducting microresonators at high magnetic fields
is thus desirable for multi-frequency ESR.  The ability to tune the resonance frequency on-demand also allows greater
control of spin dynamics \cite{kaupp_purcell-enhanced_2016,bienfait_controlling_2016} as well as the coupled spin-cavity
dynamics in the strong-coupling regime where spin ensembles are used as quantum memories
\cite{kubo_hybrid_2011,grezes_multimode_2014}.  In this work, we fabricate frequency-tunable CPW resonators that are
compatible with high magnetic fields following recently developed techniques
\cite{annunziata_tunable_2010,vissers_frequency-tunable_2015,adamyan_tunable_2016}. We show multi-frequency ESR with
$^{31}$P donors in $^{28}$Si, finding no effect on spin coherence when the resonators are operated at different
frequencies. We show that the resonance frequency can be tuned in no more than $\sim$270 ns. As a practical application,
we take advantage of the rapid tunability of our resonators to manipulate $^{31}$P and $^{75}$As donor spin ensembles
that are $33$ MHz apart in a single pulse sequence.

The resonators are fabricated from a 20 nm NbTiN thin film ($T_C = 13.8$ K) on a c-axis sapphire substrate using optical
lithography followed by reactive-ion etching with an SF$_6$/Ar plasma. Devices are wirebonded to a copper printed
circuit board equipped with microwave connectors and attached to a rotatable sample holder which enables
\textit{in situ} alignment of the device with its surface parallel to the externally applied magnetic field.

\begin{figure}
     \includegraphics{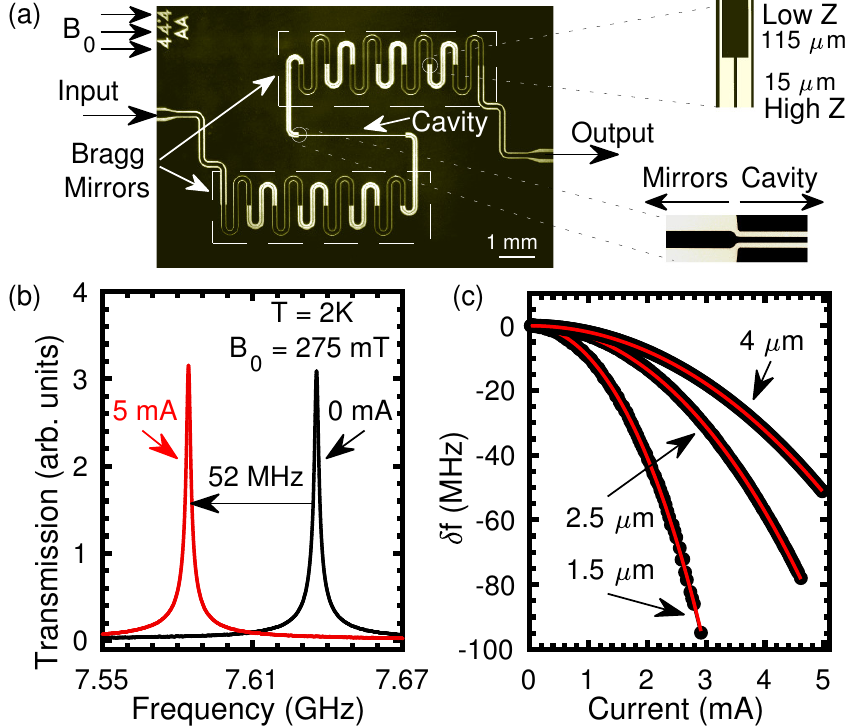}\caption{\label{fig1}(a) Optical micrograph of a tunable superconducting
         coplanar photonic bandgap microresonator. The device is patterned from a 20 nm NbTiN thin film on a c-axis sapphire
         substrate. The photonic bandgap is defined by the two Bragg mirrors on either side of the cavity. The Bragg
         mirrors are implemented using stepped impedances (top inset) and the resonance frequency is determined by the
         length of the cavity. The stepped impedances are tapered at both ends of the cavity to avoid current crowding
         (bottom inset). The center pin is continuous throughout the device from the input port to the output port.
         Three such devices are investigated with cavity center pin widths of 1.5 $\mu$m, 2.5 $\mu$m and 4 $\mu$m.
         (b) The resonance frequency of the 4 $\mu$m device shifts by $\delta$f = 52 MHz for a bias current of 5 mA,
         with no change in the quality factor of 3000.  (c) Dependence of $\delta$f on bias current for the three
         different devices. The maximum value of $\delta$f before the bias current exceeds the critical current is seen
     to be {95 MHz, 78 MHz and 52 MHz} for the {1.5 $\mu$m, 2.5 $\mu$m and 4 $\mu$m} devices, respectively. The fits
 (red lines) are generated using Eq.\ \ref{eq:fofI}.}
 \end{figure}

An optical micrograph of a device is shown in Fig.\ \ref{fig1}a. The structure is a photonic bandgap (PBG) resonator
with two Bragg mirrors on either side of a cavity which defines the resonance frequency. A key feature of the PBG
resonator is that the center pin is continuous throughout the device, enabling the application of bias currents.
Further details of the PBG resonator design have been discussed previously \cite{liu_quantum_2017, Sigillito2017}.  The
Bragg mirrors are constructed using four periods of stepped-impedance waveguides, where a period is defined as a length
of low-impedance (35 $\Omega$) followed by a length of high-impedance (137 $\Omega$) transmission line. The top inset of
Fig.\ \ref{fig1}a shows a magnified view near one of the impedance steps. Three devices are fabricated where the width
of the center pin of the cavity is $1.5$, $2.5$ and $4~\mu$m. Care is taken to prevent current crowding
\cite{Hortensius2012,clem_geometry-dependent_2011} at both ends of the cavity by tapering from the impedance steps, as
shown in the bottom inset of Fig.\ \ref{fig1}a.

Microwave transmission through each device is monitored at a temperature of 2K while a bias current is applied to the
center pin. With no bias current, the resonance frequency of the $4~\mu$m device is 7636.6 MHz (Fig.\ \ref{fig1}b) with
a temperature-limited loaded quality factor of $\sim$3000 and a coupling coefficient of 0.6. With $5$ mA of current applied to
this device, the resonance frequency shifts to 7584.3 MHz, indicating a frequency shift, $\delta$f = 52 MHz.
The device maintains its quality factor as the bias current is increased from 0 mA to 5
mA. The critical current of the device is measured to be $5.014$ mA, in agreement with previously reported values for
similar NbTiN wires where current crowding was minimized \cite{Hortensius2012}.

Fig.\ \ref{fig1}c shows the dependence of $\delta$f as a function of the applied bias current for all three devices.
The 1.5$~\mu$m device shows the largest change in resonance frequency, $\delta$f = 95 MHz, while the 2.5$~\mu$m device
shows a change of $\delta$f = 78 MHz before the applied current exceeds the critical current of the superconductor. 

The resonance frequency shift follows a well-known dependence on the applied bias current
\cite{annunziata_tunable_2010,zmuidzinas_superconducting_2012}. The total inductance of our transmission lines is
composed of both geometric and kinetic inductance. The application of a bias current, $i$, modulates the kinetic
inductance, resulting in a change from the zero-current resonance frequency, $\text{f}_0$, according to the expression
\begin{equation}\label{eq:fofI} 
    \frac{\delta \text{f}(i)}{\text{f}_0}= -\left[\left(\frac{i}{I_2^*}\right)^2 + \left(\frac{i}{I_4^*}\right)^4\right] 
\end{equation} 
where we note that the response is always negative, independent of the sign of the current
\cite{zmuidzinas_superconducting_2012}. Here, $I_2^*$ and $I_4^*$ are parameters that set the scale of the current
non-linearity of the kinetic inductance. We fit the measured values of $\delta$f for our three devices and extract
$I_2^* = \{28.9,50.1,62.5\}$ mA and $I_4^* = \{13.9,26.3,35.7\}$ mA for the $\{1.5,2.5,4\}~\mu$m devices, respectively.
The $1.5~\mu$m device shows the largest response to applied current since it has the largest kinetic inductance fraction
of the three devices. While the quartic component has typically been small in previous reports
\cite{adamyan_tunable_2016,luomahaara_kinetic_2014}, we find that it is necessary when biasing near the critical
current. Such dependence has been reported in devices with large kinetic inductance fractions
\cite{vissers_frequency-tunable_2015,kher_kinetic_2016}.

Next, we evaluate the ESR performance of our devices using a sample consisting of a 2 $\mu$m epitaxial layer of
isotopically enriched $^{28}$Si grown on high-resistivity p-type Si. The epi-layer is bulk-doped with $^{31}$P donors to
a concentration of $5\times10^{15}/$cm$^3$ and implanted with $^{75}$As donors at a nominal depth of $200$ nm to a
concentration of $5\times10^{16}$ donors/cm$^3$ as discussed in Ref.\ \citenum{Sigillito2017}.  We clip the sample on
top of the $4~\mu$m resonator with a phosphor bronze spring. We choose this resonator over the $1.5~\mu$m and $2.5~\mu$m
devices because of better expected homogeneity of the microwave magnetic field seen by donor spins in the $2~\mu$m
epi-layer.

\begin{figure}
    \includegraphics{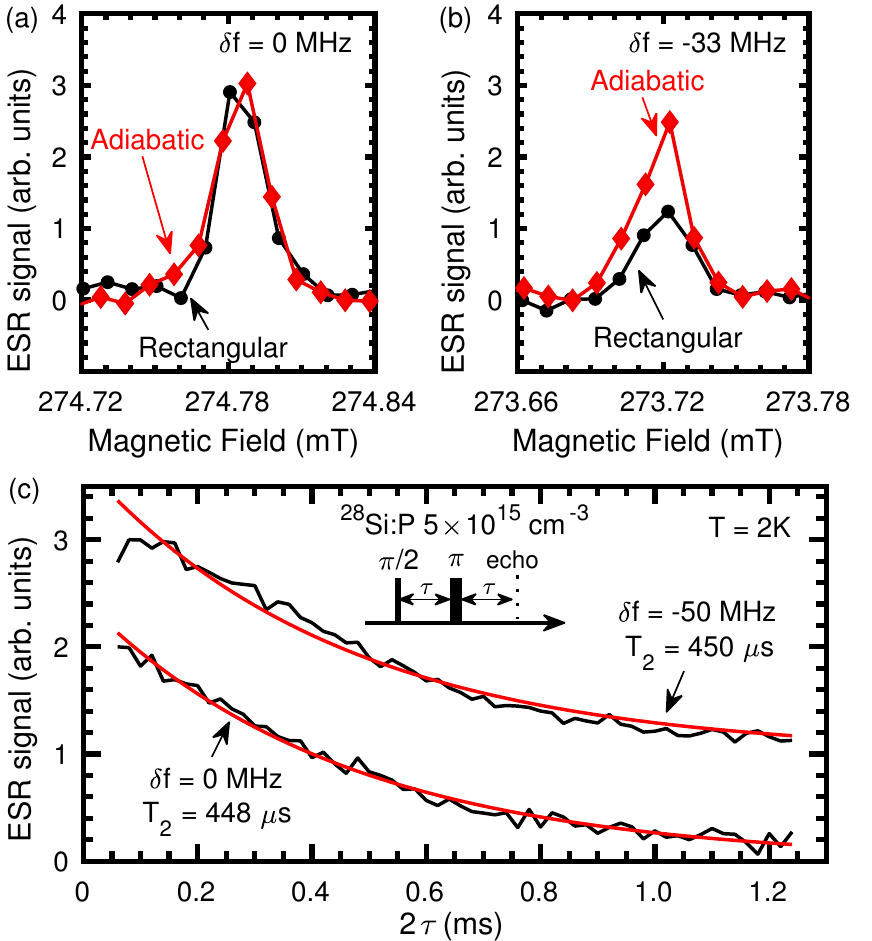} 
    \caption{\label{fig2}Echo-detected field sweeps of the $m_I = -1/2$ ESR transition of $^{31}$P donor spins in
        $^{28}$Si with 0 mA (a) or 4 mA (b) of bias current applied to the center pin of the resonator. Rectangular and
        adiabatic pulses are tested in both cases, and are found to produce similar signal intensities when no current
        flows through the center pin. On the other hand, rectangular pulses show reduced signal intensity when current
        flows through the center pin. The nonlinearity of the superconductor limits the maximum
            microwave power that can be applied at the input port of the resonator in the presence of a bias current.
            The reduced echo intensity for rectangular pulses is due to a combination of the reduced microwave power and
            magnetic field inhomogeneities introduced by the applied bias current. Adiabatic pulses are not
    power-limited and nearly recover the full signal intensity. (c) Coherence times are found to be the same with and
without bias current, indicating that the bias current noise does not introduce additional decoherence to the spins at
timescales shorter than $\sim1$ ms.} 
\end{figure}

The device and sample are cooled to 2K in a magnetic field of $B_0 = 275$ mT oriented parallel to the plane of the
superconductor. This orientation maintains the loaded quality factor of 3000 by avoiding dissipation due to the motion
of trapped flux vortices in the superconductor under the influence of microwave currents \cite{GLtheory,Hans2013}.  We
measure echo-detected field sweeps of $^{31}$P donors with and without bias currents as shown in Fig.\ \ref{fig2}. In
these measurements, the intensity of a Hahn echo ($\pi/2(\text{+x})-\tau-\pi(\text{+y})-\tau-\text{echo}$) is monitored
with $\tau = 60~\mu$s as the external magnetic field, $B_0$, is swept. Fig.\ \ref{fig2}a shows the results of field
sweeps measured without bias current, where the $m_I = -1/2$ $^{31}$P donor hyperfine line appears at $274.78$ mT. We
repeated this experiment using both rectangular ($\pi/2$-pulse $= 200$ ns, $\pi$-pulse $= 400$ ns) and adiabatic pulses
(BIR4-WURST20 \cite{Sigillito_2014}, $10~\mu$s, chirp frequency $\pm2$ MHz) with microwave powers of -29 dBm and -32
dBm, respectively, at the input port of the resonator. Adiabatic pulses have previously been shown to compensate for
rotation errors arising from microwave magnetic field inhomogeneities in the bulk-doped sample \cite{Sigillito_2014}. As
shown in Fig.\ \ref{fig2}a, the signals obtained using rectangular and adiabatic pulses are comparable, indicating good
homogeneity of the microwave magnetic field in the $2~\mu$m epi-layer.

We performed the field sweep experiment with a bias current of $i = 4$ mA applied to the center pin of the resonator
throughout the entire pulse sequence. The $^{31}$P signal is shifted to $273.72$ mT in agreement with the resonance
frequency change of 33 MHz. The center pin of the device is aligned with the external magnetic field, $B_0$, such that
the inhomogeneous magnetic field generated by the applied bias current, $B_i$, mostly adds to $B_0$ in quadrature.
However, a small misalignment ($B_0$ in the plane of NbTiN but misaligned with the long axis of the
    resonator) leads to a component of $B_i$ along $B_0$ that broadens the observed ESR transition for spins directly
    above the center pin. Such a misalignment does not affect spins above the gap between the center pin and the ground
    plane where $B_i$ remains orthogonal to $B_0$ with the misalignment. From simulations that take into account the
    full distribution of $B_i$ and the microwave magnetic field, $B_1$, we estimate that a broadening comparable to the
    ESR linewidth can be caused by a misalignment of $\sim4.7^\circ$. Additionally, we observed that microwave
    powers larger than -32 dBm affected the resonance lineshape with a bias current of 4 mA due to the non-linearity of
    the superconductor \cite{abdo_unusual_2006,swenson_operation_2013}.  The reduced signal intensity for rectangular
    pulses is due to a combination of reduced microwave power and broadening due to misalignment. By using adiabatic
pulses which are not power-limited at -32 dBm, we nearly recover the full signal intensity as shown in Fig.\
\ref{fig2}b.
   
In order to evaluate the effect of the applied bias currents on the coherence of the spin ensemble, we measure $T_2$
with and without bias currents as shown in Fig.\ \ref{fig2}c. In both cases, we use adiabatic pulses and measure the
intensity of the Hahn echo as the interpulse delay, $\tau$, is increased. We find $T_2 = 448\pm5.2~\mu$s without bias
current and $T_2 = 450\pm9.1~\mu$s with 4.9 mA bias current. The agreement of these two values indicates that there is
no effect on spin coherence due to noise in the applied bias current at timescales shorter than $\sim1$ ms. It is
conceivable that the stability of the current source will become important at longer timescales.  

\begin{figure}
     \includegraphics{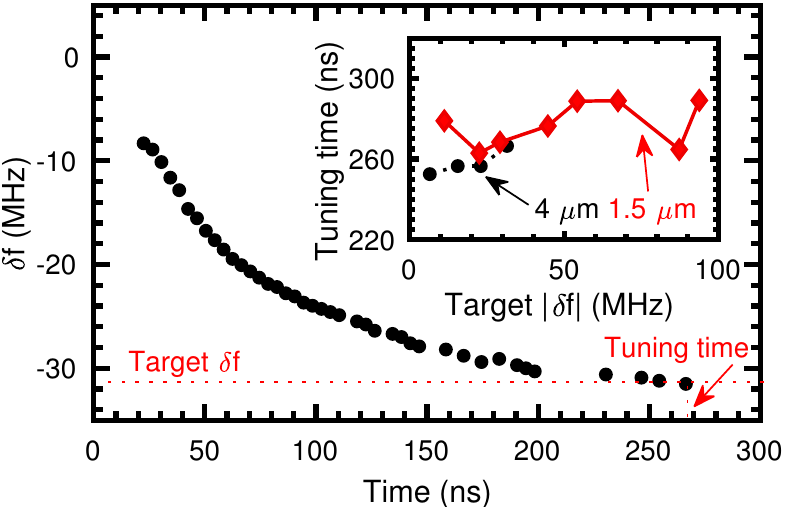}
     \caption{\label{fig3}Dynamics of the resonance frequency change in response to a 3.9 mA current pulse at time $t=0$ measured by
         monitoring microwave transmission through the device. The inset shows tuning times for various values of the target
         $\delta$f for the $4~\mu$m and $1.5~\mu$m devices. The measured tuning times are independent of the frequency
         change.}
\end{figure}

In Fig.\ \ref{fig3}, we measure how quickly we can tune the resonator to a new frequency by applying a current pulse to
the center pin and monitoring microwave transmission through the device. As the resonance frequency shifts in response
to the current pulse at $t=0$, it coincides with the frequency of the probe tone at a later time, $t = T$, enabling transmission
through the resonator and resulting in a peak in the measurement at time $T$. We show the results of this measurement for a 3.9 mA
current pulse in which the resonator reaches the target $\delta$f = 31.2 MHz in $270$ ns.  The inset compiles all
measured tuning times for the $4~\mu$m and $1.5~\mu$m devices showing no frequency dependence with an average delay of
$270$ ns. In our experiments, we used resistive power combiners and DC blocks to combine the microwaves with the bias
currents. The measured tuning times are likely limited by the time constant of our current-biasing circuit rather than
the response of the superconductor.

\begin{figure}
    \includegraphics{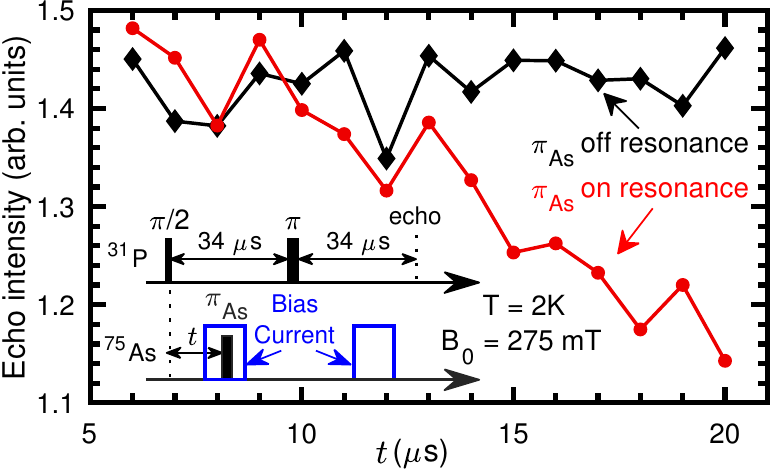} 
    \caption{\label{fig4}Double electron-electron resonance with $^{31}$P and $^{75}$As spin ensembles. The Hahn echo from
        $^{31}$P spins is monitored while a $\pi$-pulse (labeled $\pi_{\text{As}}$) is applied to the $^{75}$As spins.
        The position $t$ of this pulse is swept and its effect on the $^{31}$P spin echo is measured. The two spin
        transitions are 33 MHz apart. Details about the pulse sequence are described in the main text. When $\pi_{\text{As}}$ is
        resonant with the $^{75}$As spins, dipole-induced dephasing of the $^{31}$P spins is enhanced and the echo intensity is
        reduced by $\sim$20\%.}
\end{figure}

We demonstrate the use of this rapid tunability in a double electron-electron resonance (DEER) experiment
\cite{milov1981application,milov1984electron}. Here, we monitor the Hahn echo intensity from $^{31}$P donors while also
applying a $\pi$ pulse to $^{75}$As donors (hereafter referred to as $\pi_{\text{As}}$). The pulse sequence is shown in
the inset of Fig.\ \ref{fig4}. The effect of $\pi_{\text{As}}$ is to flip $^{75}$As donors such that the local dipolar
field which is seen by each $^{31}$P donor spin in the first free evolution period ($\tau=34~\mu$s) is different from
that of the second. This results in additional dipole-induced dephasing of the $^{31}$P donor spin ensemble due to the
random distribution of $^{75}$As donors; the effect is maximized when $\pi_{\text{As}}$ is applied close to the $^{31}$P
donor $\pi$-pulse \cite{milov_pulsed_1998}.

The $\pi_{\text{As}}$ pulse is applied at time $t$ after the $\pi/2$ pulse which tips the $^{31}$P donors. The minimum
value of $t$ is $6~\mu$s, arising from timing delays that protect the detection circuitry of our spectrometer and a $1~\mu$s
wait time for the resonator to settle to the frequency of $^{75}$As donors before the application of $\pi_\text{As}$.
The results are shown in Fig.\ \ref{fig4}, where $\pi_\text{As}$ is either on or off resonance with $^{75}$As donors,
keeping the rest of the experimental parameters constant. When $\pi_{\text{As}}$ is resonant, a reduction in the Hahn echo
intensity of $\sim20\%$ at $t=20~\mu$s is observed due to additional dephasing in the $^{31}$P donors arising from changes in
the local magnetic field environment when the $^{75}$As donor spins are flipped.

We note here that a typical fixed-frequency resonator used for DEER requires a low Q-factor of $\sim$230 in order to
accommodate the two donor frequencies that are 33 MHz apart. This is disadvantageous because the use of a low-Q
resonator requires high-power pulses to invert both donors, and also leads to poor detection sensitivity. In contrast,
our tunable resonator maintains a Q-factor of 3000 at both donor frequencies. The use of a fixed-frequency
resonator would lead to a reduction in the detected echo signal by a factor of $\sim10$ meaning an increase in averaging
time by a factor of $\sim100$ to maintain a similar signal-to-noise ratio. Therefore, our tunable resonator enables
high-sensitivity DEER with significantly reduced signal-averaging time.

We remark here on practical considerations when using these devices. (1) The large kinetic inductance non-linearity of
NbTiN results in an undesirable limitation on the maximum microwave power that can be applied for spin manipulation. For
our $4~\mu$m and $2.5~\mu$m devices, microwave powers at the input port of the resonator beyond $-15$ and $-23$ dBm
($B_1 = 0.2$ and $0.09~\mu$T), respectively, were seen to affect the resonance lineshape without 
bias currents.  This behavior has been
reported previously \cite{abdo_unusual_2006,swenson_operation_2013}, and is a signature of the strong non-linearity of
the superconductor.  (2) In our DEER experiments, we applied two bias current pulses -- one before, and another after
the $\pi$ pulse on $^{31}$P donors -- to cancel phase accumulation from the inhomogeneous magnetic fields generated by
the bias currents. We have found that bipolar current pulses can also be used to compensate for the unwanted phases
since the negative currents reverse phase accumulation due to the positive currents. Bipolar pulses are advantageous
because they only need to be applied on one side of the $\pi$ pulse. (3) Pulsed DEER experiments for biochemical samples
frequently must flip broad lines \cite{schiemann_long-range_2007}. Recent experiments have used chirped adiabatic pulses
\cite{niemeyer_broadband_2013}, but very low-Q resonators are then required, resulting in poor sensitivity. The tunable
high-Q resonator frequency could be varied during the pulse to follow the chirped pulse frequency, thus allowing broad
lines to be inverted without compromising the sensitivity of the experiment.

Our frequency-tunable devices may also be useful in experiments where a spin ensemble that is strongly coupled to the
resonator is used as a quantum memory \cite{kubo_hybrid_2011}. This enables storage of quantum information for
timescales of order $T_2^*$ of the spin ensemble \cite{brendonArXiv} and can, in principle, be extended to $T_2$, which
can exceed seconds \cite{tyryshkin_electron_2012} in silicon, by the use of refocusing $\pi$ pulses. However, the
application of a $\pi$ pulse also inverts the spin ensemble into an unstable state that can emit the stored photons into
the cavity by superradiance \cite{julsgaard_fundamental_2013,grezes_multimode_2014,brendonArXiv}. In order to avoid
uncontrolled emission of photons into the cavity, the resonator can be detuned away from the ESR frequency after the
$\pi$ pulse \cite{julsgaard_fundamental_2013}.  In this work, we have demonstrated the desired frequency tunability.
Further work is needed to elucidate the effect of magnetic field inhomogeneities introduced by the bias currents on the
coupled spin-cavity dynamics.

In summary, we have demonstrated frequency-tunable ESR microresonators fabricated from NbTiN by taking advantage of the
non-linear kinetic inductance of the superconductor. We have shown the ability to rapidly tune the resonance frequency
of these devices by as much as $\delta$f = 95 MHz within 270 ns while maintaining a quality factor of 3000 at 7.6 GHz in
a magnetic field of 275 mT at 2K. We evaluated the performance of these resonators for ESR, finding that adiabatic
(BIR4-WURST20) pulses can be used to overcome magnetic field inhomogeneities and microwave power
limitations due to the applied bias current.  We find that the applied bias currents have no effect on spin coherence
at timescales up to $1$ ms. We take advantage of the rapid frequency tunability to demonstrate high-sensitivity double
electron-electron resonance in which we address both $^{31}$P and $^{75}$As donor spin transitions that are separated by
33 MHz in a single pulse sequence. Our devices enable multi-frequency electron spin resonance with CPW microresonators
and may also be used in a hybrid quantum computation platform where spin ensembles that are strongly coupled to our
resonator serve as quantum memories.

We thank Dr.\ Robin Cantor for preparation of NbTiN thin films. Devices were fabricated in the
Princeton Institute for the Science and Technology of Materials Micro/Nano Fabrication Laboratory and the Princeton
University Quantum Device Nanofabrication Laboratory. Work at Princeton was supported by the NSF through the Princeton
MRSEC (Grant No.  DMR-01420541) and by the ARO (Grant No. W911NF-13-1-0179). Work at LBNL was performed under the
auspices of the U.S.  Department of Energy under Contract No. DE-AC02-05CH11231.


\begin{thebibliography}{30}
\makeatletter
\providecommand \@ifxundefined [1]{%
 \@ifx{#1\undefined}
}%
\providecommand \@ifnum [1]{%
 \ifnum #1\expandafter \@firstoftwo
 \else \expandafter \@secondoftwo
 \fi
}%
\providecommand \@ifx [1]{%
 \ifx #1\expandafter \@firstoftwo
 \else \expandafter \@secondoftwo
 \fi
}%
\providecommand \natexlab [1]{#1}%
\providecommand \enquote  [1]{``#1''}%
\providecommand \bibnamefont  [1]{#1}%
\providecommand \bibfnamefont [1]{#1}%
\providecommand \citenamefont [1]{#1}%
\providecommand \href@noop [0]{\@secondoftwo}%
\providecommand \href [0]{\begingroup \@sanitize@url \@href}%
\providecommand \@href[1]{\@@startlink{#1}\@@href}%
\providecommand \@@href[1]{\endgroup#1\@@endlink}%
\providecommand \@sanitize@url [0]{\catcode `\\12\catcode `\$12\catcode
  `\&12\catcode `\#12\catcode `\^12\catcode `\_12\catcode `\%12\relax}%
\providecommand \@@startlink[1]{}%
\providecommand \@@endlink[0]{}%
\providecommand \url  [0]{\begingroup\@sanitize@url \@url }%
\providecommand \@url [1]{\endgroup\@href {#1}{\urlprefix }}%
\providecommand \urlprefix  [0]{URL }%
\providecommand \Eprint [0]{\href }%
\providecommand \doibase [0]{http://dx.doi.org/}%
\providecommand \selectlanguage [0]{\@gobble}%
\providecommand \bibinfo  [0]{\@secondoftwo}%
\providecommand \bibfield  [0]{\@secondoftwo}%
\providecommand \translation [1]{[#1]}%
\providecommand \BibitemOpen [0]{}%
\providecommand \bibitemStop [0]{}%
\providecommand \bibitemNoStop [0]{.\EOS\space}%
\providecommand \EOS [0]{\spacefactor3000\relax}%
\providecommand \BibitemShut  [1]{\csname bibitem#1\endcsname}%
\let\auto@bib@innerbib\@empty
%</preamble>
\bibitem [{\citenamefont {Sigillito}\ \emph {et~al.}(2014)\citenamefont
  {Sigillito}, \citenamefont {Malissa}, \citenamefont {Tyryshkin},
  \citenamefont {Riemann}, \citenamefont {Abrosimov}, \citenamefont {Becker},
  \citenamefont {Pohl}, \citenamefont {Thewalt}, \citenamefont {Itoh},
  \citenamefont {Morton}, \citenamefont {Houck}, \citenamefont {Schuster},\
  and\ \citenamefont {Lyon}}]{Sigillito_2014}%
  \BibitemOpen
  \bibfield  {author} {\bibinfo {author} {\bibfnamefont {A.~J.}\ \bibnamefont
  {Sigillito}}, \bibinfo {author} {\bibfnamefont {H.}~\bibnamefont {Malissa}},
  \bibinfo {author} {\bibfnamefont {A.~M.}\ \bibnamefont {Tyryshkin}}, \bibinfo
  {author} {\bibfnamefont {H.}~\bibnamefont {Riemann}}, \bibinfo {author}
  {\bibfnamefont {N.~V.}\ \bibnamefont {Abrosimov}}, \bibinfo {author}
  {\bibfnamefont {P.}~\bibnamefont {Becker}}, \bibinfo {author} {\bibfnamefont
  {H.-J.}\ \bibnamefont {Pohl}}, \bibinfo {author} {\bibfnamefont {M.~L.~W.}\
  \bibnamefont {Thewalt}}, \bibinfo {author} {\bibfnamefont {K.~M.}\
  \bibnamefont {Itoh}}, \bibinfo {author} {\bibfnamefont {J.~J.~L.}\
  \bibnamefont {Morton}}, \bibinfo {author} {\bibfnamefont {A.~A.}\
  \bibnamefont {Houck}}, \bibinfo {author} {\bibfnamefont {D.~I.}\ \bibnamefont
  {Schuster}}, \ and\ \bibinfo {author} {\bibfnamefont {S.~A.}\ \bibnamefont
  {Lyon}},\ }\href {\doibase 10.1063/1.4881613} {\bibfield  {journal} {\bibinfo
   {journal} {Appl. Phys. Lett.}\ }\textbf {\bibinfo {volume} {104}},\ \bibinfo
  {pages} {222407} (\bibinfo {year} {2014})}\BibitemShut {NoStop}%
\bibitem [{\citenamefont {Sigillito}\ \emph {et~al.}(2017)\citenamefont
  {Sigillito}, \citenamefont {Tyryshkin}, \citenamefont {Schenkel},
  \citenamefont {Houck},\ and\ \citenamefont {Lyon}}]{Sigillito2017}%
  \BibitemOpen
  \bibfield  {author} {\bibinfo {author} {\bibfnamefont {A.~J.}\ \bibnamefont
  {Sigillito}}, \bibinfo {author} {\bibfnamefont {A.~M.}\ \bibnamefont
  {Tyryshkin}}, \bibinfo {author} {\bibfnamefont {T.}~\bibnamefont {Schenkel}},
  \bibinfo {author} {\bibfnamefont {A.~A.}\ \bibnamefont {Houck}}, \ and\
  \bibinfo {author} {\bibfnamefont {S.~A.}\ \bibnamefont {Lyon}},\ }\href
  {http://arxiv.org/pdf/1701.06650.pdf} {\bibfield  {journal} {\bibinfo
  {journal} {arXiv:1701.06650 [quant-ph]}\ } (\bibinfo {year}
  {2017})}\BibitemShut {NoStop}%
\bibitem [{\citenamefont {Eichler}\ \emph {et~al.}(2017)\citenamefont
  {Eichler}, \citenamefont {Sigillito}, \citenamefont {Lyon},\ and\
  \citenamefont {Petta}}]{eichler_electron_2017}%
  \BibitemOpen
  \bibfield  {author} {\bibinfo {author} {\bibfnamefont {C.}~\bibnamefont
  {Eichler}}, \bibinfo {author} {\bibfnamefont {A.~J.}\ \bibnamefont
  {Sigillito}}, \bibinfo {author} {\bibfnamefont {S.~A.}\ \bibnamefont {Lyon}},
  \ and\ \bibinfo {author} {\bibfnamefont {J.~R.}\ \bibnamefont {Petta}},\
  }\href {\doibase 10.1103/PhysRevLett.118.037701} {\bibfield  {journal}
  {\bibinfo  {journal} {Phys. Rev. Lett.}\ }\textbf {\bibinfo {volume} {118}},\
  \bibinfo {pages} {037701} (\bibinfo {year} {2017})}\BibitemShut {NoStop}%
\bibitem [{\citenamefont {Bienfait}\ \emph
  {et~al.}(2016{\natexlab{a}})\citenamefont {Bienfait}, \citenamefont {Pla},
  \citenamefont {Kubo}, \citenamefont {Stern}, \citenamefont {Zhou},
  \citenamefont {Lo}, \citenamefont {Weis}, \citenamefont {Schenkel},
  \citenamefont {Thewalt}, \citenamefont {Vion}, \citenamefont {Esteve},
  \citenamefont {Julsgaard}, \citenamefont {M{\o}lmer}, \citenamefont
  {Morton},\ and\ \citenamefont {Bertet}}]{bienfait_reaching_2016}%
  \BibitemOpen
  \bibfield  {author} {\bibinfo {author} {\bibfnamefont {A.}~\bibnamefont
  {Bienfait}}, \bibinfo {author} {\bibfnamefont {J.~J.}\ \bibnamefont {Pla}},
  \bibinfo {author} {\bibfnamefont {Y.}~\bibnamefont {Kubo}}, \bibinfo {author}
  {\bibfnamefont {M.}~\bibnamefont {Stern}}, \bibinfo {author} {\bibfnamefont
  {X.}~\bibnamefont {Zhou}}, \bibinfo {author} {\bibfnamefont {C.~C.}\
  \bibnamefont {Lo}}, \bibinfo {author} {\bibfnamefont {C.~D.}\ \bibnamefont
  {Weis}}, \bibinfo {author} {\bibfnamefont {T.}~\bibnamefont {Schenkel}},
  \bibinfo {author} {\bibfnamefont {M.~L.~W.}\ \bibnamefont {Thewalt}},
  \bibinfo {author} {\bibfnamefont {D.}~\bibnamefont {Vion}}, \bibinfo {author}
  {\bibfnamefont {D.}~\bibnamefont {Esteve}}, \bibinfo {author} {\bibfnamefont
  {B.}~\bibnamefont {Julsgaard}}, \bibinfo {author} {\bibfnamefont
  {K.}~\bibnamefont {M{\o}lmer}}, \bibinfo {author} {\bibfnamefont {J.~J.~L.}\
  \bibnamefont {Morton}}, \ and\ \bibinfo {author} {\bibfnamefont
  {P.}~\bibnamefont {Bertet}},\ }\href {\doibase 10.1038/nnano.2015.282}
  {\bibfield  {journal} {\bibinfo  {journal} {Nat. Nanotechnol.}\ }\textbf
  {\bibinfo {volume} {11}},\ \bibinfo {pages} {253} (\bibinfo {year}
  {2016}{\natexlab{a}})}\BibitemShut {NoStop}%
\bibitem [{\citenamefont {Petersson}\ \emph {et~al.}(2012)\citenamefont
  {Petersson}, \citenamefont {McFaul}, \citenamefont {Schroer}, \citenamefont
  {Jung}, \citenamefont {Taylor}, \citenamefont {Houck},\ and\ \citenamefont
  {Petta}}]{petersson_circuit_2012}%
  \BibitemOpen
  \bibfield  {author} {\bibinfo {author} {\bibfnamefont {K.~D.}\ \bibnamefont
  {Petersson}}, \bibinfo {author} {\bibfnamefont {L.~W.}\ \bibnamefont
  {McFaul}}, \bibinfo {author} {\bibfnamefont {M.~D.}\ \bibnamefont {Schroer}},
  \bibinfo {author} {\bibfnamefont {M.}~\bibnamefont {Jung}}, \bibinfo {author}
  {\bibfnamefont {J.~M.}\ \bibnamefont {Taylor}}, \bibinfo {author}
  {\bibfnamefont {A.~A.}\ \bibnamefont {Houck}}, \ and\ \bibinfo {author}
  {\bibfnamefont {J.~R.}\ \bibnamefont {Petta}},\ }\href {\doibase
  10.1038/nature11559} {\bibfield  {journal} {\bibinfo  {journal} {Nature}\
  }\textbf {\bibinfo {volume} {490}},\ \bibinfo {pages} {380} (\bibinfo {year}
  {2012})}\BibitemShut {NoStop}%
\bibitem [{\citenamefont {Kubo}\ \emph {et~al.}(2011)\citenamefont {Kubo},
  \citenamefont {Grezes}, \citenamefont {Dewes}, \citenamefont {Umeda},
  \citenamefont {Isoya}, \citenamefont {Sumiya}, \citenamefont {Morishita},
  \citenamefont {Abe}, \citenamefont {Onoda}, \citenamefont {Ohshima},
  \citenamefont {Jacques}, \citenamefont {Dr{\'e}au}, \citenamefont {Roch},
  \citenamefont {Diniz}, \citenamefont {Auffeves}, \citenamefont {Vion},
  \citenamefont {Esteve},\ and\ \citenamefont {Bertet}}]{kubo_hybrid_2011}%
  \BibitemOpen
  \bibfield  {author} {\bibinfo {author} {\bibfnamefont {Y.}~\bibnamefont
  {Kubo}}, \bibinfo {author} {\bibfnamefont {C.}~\bibnamefont {Grezes}},
  \bibinfo {author} {\bibfnamefont {A.}~\bibnamefont {Dewes}}, \bibinfo
  {author} {\bibfnamefont {T.}~\bibnamefont {Umeda}}, \bibinfo {author}
  {\bibfnamefont {J.}~\bibnamefont {Isoya}}, \bibinfo {author} {\bibfnamefont
  {H.}~\bibnamefont {Sumiya}}, \bibinfo {author} {\bibfnamefont
  {N.}~\bibnamefont {Morishita}}, \bibinfo {author} {\bibfnamefont
  {H.}~\bibnamefont {Abe}}, \bibinfo {author} {\bibfnamefont {S.}~\bibnamefont
  {Onoda}}, \bibinfo {author} {\bibfnamefont {T.}~\bibnamefont {Ohshima}},
  \bibinfo {author} {\bibfnamefont {V.}~\bibnamefont {Jacques}}, \bibinfo
  {author} {\bibfnamefont {A.}~\bibnamefont {Dr{\'e}au}}, \bibinfo {author}
  {\bibfnamefont {J.-F.}\ \bibnamefont {Roch}}, \bibinfo {author}
  {\bibfnamefont {I.}~\bibnamefont {Diniz}}, \bibinfo {author} {\bibfnamefont
  {A.}~\bibnamefont {Auffeves}}, \bibinfo {author} {\bibfnamefont
  {D.}~\bibnamefont {Vion}}, \bibinfo {author} {\bibfnamefont {D.}~\bibnamefont
  {Esteve}}, \ and\ \bibinfo {author} {\bibfnamefont {P.}~\bibnamefont
  {Bertet}},\ }\href {\doibase 10.1103/PhysRevLett.107.220501} {\bibfield
  {journal} {\bibinfo  {journal} {Phys. Rev. Lett.}\ }\textbf {\bibinfo
  {volume} {107}},\ \bibinfo {pages} {220501} (\bibinfo {year}
  {2011})}\BibitemShut {NoStop}%
\bibitem [{\citenamefont {Kaupp}\ \emph {et~al.}(2016)\citenamefont {Kaupp},
  \citenamefont {H{\"u}mmer}, \citenamefont {Mader}, \citenamefont
  {Schlederer}, \citenamefont {Benedikter}, \citenamefont {Haeusser},
  \citenamefont {Chang}, \citenamefont {Fedder}, \citenamefont {H\"ansch},\
  and\ \citenamefont {Hunger}}]{kaupp_purcell-enhanced_2016}%
  \BibitemOpen
  \bibfield  {author} {\bibinfo {author} {\bibfnamefont {H.}~\bibnamefont
  {Kaupp}}, \bibinfo {author} {\bibfnamefont {T.}~\bibnamefont {H{\"u}mmer}},
  \bibinfo {author} {\bibfnamefont {M.}~\bibnamefont {Mader}}, \bibinfo
  {author} {\bibfnamefont {B.}~\bibnamefont {Schlederer}}, \bibinfo {author}
  {\bibfnamefont {J.}~\bibnamefont {Benedikter}}, \bibinfo {author}
  {\bibfnamefont {P.}~\bibnamefont {Haeusser}}, \bibinfo {author}
  {\bibfnamefont {H.-C.}\ \bibnamefont {Chang}}, \bibinfo {author}
  {\bibfnamefont {H.}~\bibnamefont {Fedder}}, \bibinfo {author} {\bibfnamefont
  {T.~W.}\ \bibnamefont {H\"ansch}}, \ and\ \bibinfo {author} {\bibfnamefont
  {D.}~\bibnamefont {Hunger}},\ }\href {\doibase
  10.1103/PhysRevApplied.6.054010} {\bibfield  {journal} {\bibinfo  {journal}
  {Phys. Rev. Applied}\ }\textbf {\bibinfo {volume} {6}},\ \bibinfo {pages}
  {054010} (\bibinfo {year} {2016})}\BibitemShut {NoStop}%
\bibitem [{\citenamefont {Bienfait}\ \emph
  {et~al.}(2016{\natexlab{b}})\citenamefont {Bienfait}, \citenamefont {Pla},
  \citenamefont {Kubo}, \citenamefont {Zhou}, \citenamefont {Stern},
  \citenamefont {Lo}, \citenamefont {Weis}, \citenamefont {Schenkel},
  \citenamefont {Vion}, \citenamefont {Esteve}, \citenamefont {Morton},\ and\
  \citenamefont {Bertet}}]{bienfait_controlling_2016}%
  \BibitemOpen
  \bibfield  {author} {\bibinfo {author} {\bibfnamefont {A.}~\bibnamefont
  {Bienfait}}, \bibinfo {author} {\bibfnamefont {J.~J.}\ \bibnamefont {Pla}},
  \bibinfo {author} {\bibfnamefont {Y.}~\bibnamefont {Kubo}}, \bibinfo {author}
  {\bibfnamefont {X.}~\bibnamefont {Zhou}}, \bibinfo {author} {\bibfnamefont
  {M.}~\bibnamefont {Stern}}, \bibinfo {author} {\bibfnamefont {C.~C.}\
  \bibnamefont {Lo}}, \bibinfo {author} {\bibfnamefont {C.~D.}\ \bibnamefont
  {Weis}}, \bibinfo {author} {\bibfnamefont {T.}~\bibnamefont {Schenkel}},
  \bibinfo {author} {\bibfnamefont {D.}~\bibnamefont {Vion}}, \bibinfo {author}
  {\bibfnamefont {D.}~\bibnamefont {Esteve}}, \bibinfo {author} {\bibfnamefont
  {J.~J.~L.}\ \bibnamefont {Morton}}, \ and\ \bibinfo {author} {\bibfnamefont
  {P.}~\bibnamefont {Bertet}},\ }\href {\doibase 10.1038/nature16944}
  {\bibfield  {journal} {\bibinfo  {journal} {Nature}\ }\textbf {\bibinfo
  {volume} {531}},\ \bibinfo {pages} {74} (\bibinfo {year}
  {2016}{\natexlab{b}})}\BibitemShut {NoStop}%
\bibitem [{\citenamefont {Grezes}\ \emph {et~al.}(2014)\citenamefont {Grezes},
  \citenamefont {Julsgaard}, \citenamefont {Kubo}, \citenamefont {Stern},
  \citenamefont {Umeda}, \citenamefont {Isoya}, \citenamefont {Sumiya},
  \citenamefont {Abe}, \citenamefont {Onoda}, \citenamefont {Ohshima},
  \citenamefont {Jacques}, \citenamefont {Esteve}, \citenamefont {Vion},
  \citenamefont {Esteve}, \citenamefont {M{\o}lmer},\ and\ \citenamefont
  {Bertet}}]{grezes_multimode_2014}%
  \BibitemOpen
  \bibfield  {author} {\bibinfo {author} {\bibfnamefont {C.}~\bibnamefont
  {Grezes}}, \bibinfo {author} {\bibfnamefont {B.}~\bibnamefont {Julsgaard}},
  \bibinfo {author} {\bibfnamefont {Y.}~\bibnamefont {Kubo}}, \bibinfo {author}
  {\bibfnamefont {M.}~\bibnamefont {Stern}}, \bibinfo {author} {\bibfnamefont
  {T.}~\bibnamefont {Umeda}}, \bibinfo {author} {\bibfnamefont
  {J.}~\bibnamefont {Isoya}}, \bibinfo {author} {\bibfnamefont
  {H.}~\bibnamefont {Sumiya}}, \bibinfo {author} {\bibfnamefont
  {H.}~\bibnamefont {Abe}}, \bibinfo {author} {\bibfnamefont {S.}~\bibnamefont
  {Onoda}}, \bibinfo {author} {\bibfnamefont {T.}~\bibnamefont {Ohshima}},
  \bibinfo {author} {\bibfnamefont {V.}~\bibnamefont {Jacques}}, \bibinfo
  {author} {\bibfnamefont {J.}~\bibnamefont {Esteve}}, \bibinfo {author}
  {\bibfnamefont {D.}~\bibnamefont {Vion}}, \bibinfo {author} {\bibfnamefont
  {D.}~\bibnamefont {Esteve}}, \bibinfo {author} {\bibfnamefont
  {K.}~\bibnamefont {M{\o}lmer}}, \ and\ \bibinfo {author} {\bibfnamefont
  {P.}~\bibnamefont {Bertet}},\ }\href {\doibase 10.1103/PhysRevX.4.021049}
  {\bibfield  {journal} {\bibinfo  {journal} {Phys. Rev. X}\ }\textbf {\bibinfo
  {volume} {4}},\ \bibinfo {pages} {021049} (\bibinfo {year}
  {2014})}\BibitemShut {NoStop}%
\bibitem [{\citenamefont {Annunziata}\ \emph {et~al.}(2010)\citenamefont
  {Annunziata}, \citenamefont {Santavicca}, \citenamefont {Frunzio},
  \citenamefont {Catelani}, \citenamefont {Rooks}, \citenamefont {Frydman},\
  and\ \citenamefont {Prober}}]{annunziata_tunable_2010}%
  \BibitemOpen
  \bibfield  {author} {\bibinfo {author} {\bibfnamefont {A.~J.}\ \bibnamefont
  {Annunziata}}, \bibinfo {author} {\bibfnamefont {D.~F.}\ \bibnamefont
  {Santavicca}}, \bibinfo {author} {\bibfnamefont {L.}~\bibnamefont {Frunzio}},
  \bibinfo {author} {\bibfnamefont {G.}~\bibnamefont {Catelani}}, \bibinfo
  {author} {\bibfnamefont {M.~J.}\ \bibnamefont {Rooks}}, \bibinfo {author}
  {\bibfnamefont {A.}~\bibnamefont {Frydman}}, \ and\ \bibinfo {author}
  {\bibfnamefont {D.~E.}\ \bibnamefont {Prober}},\ }\href {\doibase
  10.1088/0957-4484/21/44/445202} {\bibfield  {journal} {\bibinfo  {journal}
  {Nanotechnology}\ }\textbf {\bibinfo {volume} {21}},\ \bibinfo {pages}
  {445202} (\bibinfo {year} {2010})}\BibitemShut {NoStop}%
\bibitem [{\citenamefont {Vissers}\ \emph {et~al.}(2015)\citenamefont
  {Vissers}, \citenamefont {Hubmayr}, \citenamefont {Sandberg}, \citenamefont
  {Chaudhuri}, \citenamefont {Bockstiegel},\ and\ \citenamefont
  {Gao}}]{vissers_frequency-tunable_2015}%
  \BibitemOpen
  \bibfield  {author} {\bibinfo {author} {\bibfnamefont {M.~R.}\ \bibnamefont
  {Vissers}}, \bibinfo {author} {\bibfnamefont {J.}~\bibnamefont {Hubmayr}},
  \bibinfo {author} {\bibfnamefont {M.}~\bibnamefont {Sandberg}}, \bibinfo
  {author} {\bibfnamefont {S.}~\bibnamefont {Chaudhuri}}, \bibinfo {author}
  {\bibfnamefont {C.}~\bibnamefont {Bockstiegel}}, \ and\ \bibinfo {author}
  {\bibfnamefont {J.}~\bibnamefont {Gao}},\ }\href {\doibase 10.1063/1.4927444}
  {\bibfield  {journal} {\bibinfo  {journal} {Appl. Phys. Lett.}\ }\textbf
  {\bibinfo {volume} {107}},\ \bibinfo {pages} {062601} (\bibinfo {year}
  {2015})}\BibitemShut {NoStop}%
\bibitem [{\citenamefont {Adamyan}, \citenamefont {Kubatkin},\ and\
  \citenamefont {Danilov}(2016)}]{adamyan_tunable_2016}%
  \BibitemOpen
  \bibfield  {author} {\bibinfo {author} {\bibfnamefont {A.~A.}\ \bibnamefont
  {Adamyan}}, \bibinfo {author} {\bibfnamefont {S.~E.}\ \bibnamefont
  {Kubatkin}}, \ and\ \bibinfo {author} {\bibfnamefont {A.~V.}\ \bibnamefont
  {Danilov}},\ }\href {\doibase 10.1063/1.4947579} {\bibfield  {journal}
  {\bibinfo  {journal} {Appl. Phys. Lett.}\ }\textbf {\bibinfo {volume}
  {108}},\ \bibinfo {pages} {172601} (\bibinfo {year} {2016})}\BibitemShut
  {NoStop}%
\bibitem [{\citenamefont {Liu}\ and\ \citenamefont
  {Houck}(2017)}]{liu_quantum_2017}%
  \BibitemOpen
  \bibfield  {author} {\bibinfo {author} {\bibfnamefont {Y.}~\bibnamefont
  {Liu}}\ and\ \bibinfo {author} {\bibfnamefont {A.~A.}\ \bibnamefont
  {Houck}},\ }\href {\doibase 10.1038/nphys3834} {\bibfield  {journal}
  {\bibinfo  {journal} {Nature Phys.}\ }\textbf {\bibinfo {volume} {13}},\
  \bibinfo {pages} {48} (\bibinfo {year} {2017})}\BibitemShut {NoStop}%
\bibitem [{\citenamefont {Hortensius}\ \emph {et~al.}(2012)\citenamefont
  {Hortensius}, \citenamefont {Driessen}, \citenamefont {Klapwijk},
  \citenamefont {Berggren},\ and\ \citenamefont {Clem}}]{Hortensius2012}%
  \BibitemOpen
  \bibfield  {author} {\bibinfo {author} {\bibfnamefont {H.~L.}\ \bibnamefont
  {Hortensius}}, \bibinfo {author} {\bibfnamefont {E.~F.~C.}\ \bibnamefont
  {Driessen}}, \bibinfo {author} {\bibfnamefont {T.~M.}\ \bibnamefont
  {Klapwijk}}, \bibinfo {author} {\bibfnamefont {K.~K.}\ \bibnamefont
  {Berggren}}, \ and\ \bibinfo {author} {\bibfnamefont {J.~R.}\ \bibnamefont
  {Clem}},\ }\href {\doibase 10.1063/1.4711217} {\bibfield  {journal} {\bibinfo
   {journal} {Appl. Phys. Lett.}\ }\textbf {\bibinfo {volume} {100}},\ \bibinfo
  {pages} {182602} (\bibinfo {year} {2012})},\ \Eprint
  {http://arxiv.org/abs/http://dx.doi.org/10.1063/1.4711217}
  {http://dx.doi.org/10.1063/1.4711217} \BibitemShut {NoStop}%
\bibitem [{\citenamefont {Clem}\ and\ \citenamefont
  {Berggren}(2011)}]{clem_geometry-dependent_2011}%
  \BibitemOpen
  \bibfield  {author} {\bibinfo {author} {\bibfnamefont {J.~R.}\ \bibnamefont
  {Clem}}\ and\ \bibinfo {author} {\bibfnamefont {K.~K.}\ \bibnamefont
  {Berggren}},\ }\href {\doibase 10.1103/PhysRevB.84.174510} {\bibfield
  {journal} {\bibinfo  {journal} {Phys. Rev. B}\ }\textbf {\bibinfo {volume}
  {84}},\ \bibinfo {pages} {174510} (\bibinfo {year} {2011})}\BibitemShut
  {NoStop}%
\bibitem [{\citenamefont {Zmuidzinas}(2012)}]{zmuidzinas_superconducting_2012}%
  \BibitemOpen
  \bibfield  {author} {\bibinfo {author} {\bibfnamefont {J.}~\bibnamefont
  {Zmuidzinas}},\ }\href {\doibase 10.1146/annurev-conmatphys-020911-125022}
  {\bibfield  {journal} {\bibinfo  {journal} {Annu. Rev. Condens. Matter
  Phys.}\ }\textbf {\bibinfo {volume} {3}},\ \bibinfo {pages} {169} (\bibinfo
  {year} {2012})}\BibitemShut {NoStop}%
\bibitem [{\citenamefont {Luomahaara}\ \emph {et~al.}(2014)\citenamefont
  {Luomahaara}, \citenamefont {Vesterinen}, \citenamefont {Gr{\"o}nberg},\ and\
  \citenamefont {Hassel}}]{luomahaara_kinetic_2014}%
  \BibitemOpen
  \bibfield  {author} {\bibinfo {author} {\bibfnamefont {J.}~\bibnamefont
  {Luomahaara}}, \bibinfo {author} {\bibfnamefont {V.}~\bibnamefont
  {Vesterinen}}, \bibinfo {author} {\bibfnamefont {L.}~\bibnamefont
  {Gr{\"o}nberg}}, \ and\ \bibinfo {author} {\bibfnamefont {J.}~\bibnamefont
  {Hassel}},\ }\href {\doibase 10.1038/ncomms5872} {\bibfield  {journal}
  {\bibinfo  {journal} {Nat. Commun.}\ }\textbf {\bibinfo {volume} {5}},\
  \bibinfo {pages} {4872} (\bibinfo {year} {2014})}\BibitemShut {NoStop}%
\bibitem [{\citenamefont {Kher}\ \emph {et~al.}(2016)\citenamefont {Kher},
  \citenamefont {Day}, \citenamefont {Eom}, \citenamefont {Zmuidzinas},\ and\
  \citenamefont {Leduc}}]{kher_kinetic_2016}%
  \BibitemOpen
  \bibfield  {author} {\bibinfo {author} {\bibfnamefont {A.}~\bibnamefont
  {Kher}}, \bibinfo {author} {\bibfnamefont {P.~K.}\ \bibnamefont {Day}},
  \bibinfo {author} {\bibfnamefont {B.~H.}\ \bibnamefont {Eom}}, \bibinfo
  {author} {\bibfnamefont {J.}~\bibnamefont {Zmuidzinas}}, \ and\ \bibinfo
  {author} {\bibfnamefont {H.~G.}\ \bibnamefont {Leduc}},\ }\href {\doibase
  10.1007/s10909-015-1364-0} {\bibfield  {journal} {\bibinfo  {journal} {J. Low
  Temp. Phys.}\ }\textbf {\bibinfo {volume} {184}},\ \bibinfo {pages} {480}
  (\bibinfo {year} {2016})}\BibitemShut {NoStop}%
\bibitem [{\citenamefont {Ginzburg}\ and\ \citenamefont
  {Landau}(1950)}]{GLtheory}%
  \BibitemOpen
  \bibfield  {author} {\bibinfo {author} {\bibfnamefont {V.}~\bibnamefont
  {Ginzburg}}\ and\ \bibinfo {author} {\bibfnamefont {L.}~\bibnamefont
  {Landau}},\ }\href {www.scopus.com} {\bibfield  {journal} {\bibinfo
  {journal} {Zh. Eksperiment. Teoret. Fiz.}\ }\textbf {\bibinfo {volume}
  {20}},\ \bibinfo {pages} {1064} (\bibinfo {year} {1950})}\BibitemShut
  {NoStop}%
\bibitem [{\citenamefont {Malissa}\ \emph {et~al.}(2013)\citenamefont
  {Malissa}, \citenamefont {Schuster}, \citenamefont {Tyryshkin}, \citenamefont
  {Houck},\ and\ \citenamefont {Lyon}}]{Hans2013}%
  \BibitemOpen
  \bibfield  {author} {\bibinfo {author} {\bibfnamefont {H.}~\bibnamefont
  {Malissa}}, \bibinfo {author} {\bibfnamefont {D.~I.}\ \bibnamefont
  {Schuster}}, \bibinfo {author} {\bibfnamefont {A.~M.}\ \bibnamefont
  {Tyryshkin}}, \bibinfo {author} {\bibfnamefont {A.~A.}\ \bibnamefont
  {Houck}}, \ and\ \bibinfo {author} {\bibfnamefont {S.~A.}\ \bibnamefont
  {Lyon}},\ }\href {\doibase 10.1063/1.4792205} {\bibfield  {journal} {\bibinfo
   {journal} {Rev. Sci. Instrum.}\ }\textbf {\bibinfo {volume} {84}},\ \bibinfo
  {pages} {025116} (\bibinfo {year} {2013})}\BibitemShut {NoStop}%
\bibitem [{\citenamefont {Abdo}\ \emph {et~al.}(2006)\citenamefont {Abdo},
  \citenamefont {Segev}, \citenamefont {Shtempluck},\ and\ \citenamefont
  {Buks}}]{abdo_unusual_2006}%
  \BibitemOpen
  \bibfield  {author} {\bibinfo {author} {\bibfnamefont {B.}~\bibnamefont
  {Abdo}}, \bibinfo {author} {\bibfnamefont {E.}~\bibnamefont {Segev}},
  \bibinfo {author} {\bibfnamefont {O.}~\bibnamefont {Shtempluck}}, \ and\
  \bibinfo {author} {\bibfnamefont {E.}~\bibnamefont {Buks}},\ }\href {\doibase
  10.1088/1742-6596/43/1/329} {\bibfield  {journal} {\bibinfo  {journal} {J.
  Phys.: Conf. Ser.}\ }\textbf {\bibinfo {volume} {43}},\ \bibinfo {pages}
  {1346} (\bibinfo {year} {2006})}\BibitemShut {NoStop}%
\bibitem [{\citenamefont {Swenson}\ \emph {et~al.}(2013)\citenamefont
  {Swenson}, \citenamefont {Day}, \citenamefont {Eom}, \citenamefont {Leduc},
  \citenamefont {Llombart}, \citenamefont {McKenney}, \citenamefont
  {Noroozian},\ and\ \citenamefont {Zmuidzinas}}]{swenson_operation_2013}%
  \BibitemOpen
  \bibfield  {author} {\bibinfo {author} {\bibfnamefont {L.~J.}\ \bibnamefont
  {Swenson}}, \bibinfo {author} {\bibfnamefont {P.~K.}\ \bibnamefont {Day}},
  \bibinfo {author} {\bibfnamefont {B.~H.}\ \bibnamefont {Eom}}, \bibinfo
  {author} {\bibfnamefont {H.~G.}\ \bibnamefont {Leduc}}, \bibinfo {author}
  {\bibfnamefont {N.}~\bibnamefont {Llombart}}, \bibinfo {author}
  {\bibfnamefont {C.~M.}\ \bibnamefont {McKenney}}, \bibinfo {author}
  {\bibfnamefont {O.}~\bibnamefont {Noroozian}}, \ and\ \bibinfo {author}
  {\bibfnamefont {J.}~\bibnamefont {Zmuidzinas}},\ }\href {\doibase
  10.1063/1.4794808} {\bibfield  {journal} {\bibinfo  {journal} {J. Appl.
  Phys.}\ }\textbf {\bibinfo {volume} {113}},\ \bibinfo {pages} {104501}
  (\bibinfo {year} {2013})}\BibitemShut {NoStop}%
\bibitem [{\citenamefont {Milov}, \citenamefont {Salikhov},\ and\ \citenamefont
  {Shirov}(1981)}]{milov1981application}%
  \BibitemOpen
  \bibfield  {author} {\bibinfo {author} {\bibfnamefont {A.}~\bibnamefont
  {Milov}}, \bibinfo {author} {\bibfnamefont {K.}~\bibnamefont {Salikhov}}, \
  and\ \bibinfo {author} {\bibfnamefont {M.}~\bibnamefont {Shirov}},\
  }\href@noop {} {\bibfield  {journal} {\bibinfo  {journal} {Fiz. Tverd. Tela}\
  }\textbf {\bibinfo {volume} {23}},\ \bibinfo {pages} {975} (\bibinfo {year}
  {1981})}\BibitemShut {NoStop}%
\bibitem [{\citenamefont {Milov}, \citenamefont {Ponomarev},\ and\
  \citenamefont {Tsvetkov}(1984)}]{milov1984electron}%
  \BibitemOpen
  \bibfield  {author} {\bibinfo {author} {\bibfnamefont {A.}~\bibnamefont
  {Milov}}, \bibinfo {author} {\bibfnamefont {A.}~\bibnamefont {Ponomarev}}, \
  and\ \bibinfo {author} {\bibfnamefont {Y.~D.}\ \bibnamefont {Tsvetkov}},\
  }\href@noop {} {\bibfield  {journal} {\bibinfo  {journal} {Chem. Phys.
  Lett.}\ }\textbf {\bibinfo {volume} {110}},\ \bibinfo {pages} {67} (\bibinfo
  {year} {1984})}\BibitemShut {NoStop}%
\bibitem [{\citenamefont {Milov}, \citenamefont {Maryasov},\ and\ \citenamefont
  {Tsvetkov}(1998)}]{milov_pulsed_1998}%
  \BibitemOpen
  \bibfield  {author} {\bibinfo {author} {\bibfnamefont {A.~D.}\ \bibnamefont
  {Milov}}, \bibinfo {author} {\bibfnamefont {A.~G.}\ \bibnamefont {Maryasov}},
  \ and\ \bibinfo {author} {\bibfnamefont {Y.~D.}\ \bibnamefont {Tsvetkov}},\
  }\href {\doibase 10.1007/BF03161886} {\bibfield  {journal} {\bibinfo
  {journal} {Appl. Magn. Reson.}\ }\textbf {\bibinfo {volume} {15}},\ \bibinfo
  {pages} {107} (\bibinfo {year} {1998})}\BibitemShut {NoStop}%
\bibitem [{\citenamefont {Schiemann}\ and\ \citenamefont
  {Prisner}(2007)}]{schiemann_long-range_2007}%
  \BibitemOpen
  \bibfield  {author} {\bibinfo {author} {\bibfnamefont {O.}~\bibnamefont
  {Schiemann}}\ and\ \bibinfo {author} {\bibfnamefont {T.~F.}\ \bibnamefont
  {Prisner}},\ }\href {\doibase 10.1017/S003358350700460X} {\bibfield
  {journal} {\bibinfo  {journal} {Q Rev Biophys.}\ }\textbf {\bibinfo {volume}
  {40}},\ \bibinfo {pages} {1} (\bibinfo {year} {2007})}\BibitemShut {NoStop}%
\bibitem [{\citenamefont {Niemeyer}\ \emph {et~al.}(2013)\citenamefont
  {Niemeyer}, \citenamefont {Shim}, \citenamefont {Zhang}, \citenamefont
  {Suter}, \citenamefont {Taniguchi}, \citenamefont {Teraji}, \citenamefont
  {Abe}, \citenamefont {Onoda}, \citenamefont {Yamamoto}, \citenamefont
  {Ohshima}, \citenamefont {{J Isoya}},\ and\ \citenamefont
  {Jelezko}}]{niemeyer_broadband_2013}%
  \BibitemOpen
  \bibfield  {author} {\bibinfo {author} {\bibfnamefont {I.}~\bibnamefont
  {Niemeyer}}, \bibinfo {author} {\bibfnamefont {J.~H.}\ \bibnamefont {Shim}},
  \bibinfo {author} {\bibfnamefont {J.}~\bibnamefont {Zhang}}, \bibinfo
  {author} {\bibfnamefont {D.}~\bibnamefont {Suter}}, \bibinfo {author}
  {\bibfnamefont {T.}~\bibnamefont {Taniguchi}}, \bibinfo {author}
  {\bibfnamefont {T.}~\bibnamefont {Teraji}}, \bibinfo {author} {\bibfnamefont
  {H.}~\bibnamefont {Abe}}, \bibinfo {author} {\bibfnamefont {S.}~\bibnamefont
  {Onoda}}, \bibinfo {author} {\bibfnamefont {T.}~\bibnamefont {Yamamoto}},
  \bibinfo {author} {\bibfnamefont {T.}~\bibnamefont {Ohshima}}, \bibinfo
  {author} {\bibnamefont {{J Isoya}}}, \ and\ \bibinfo {author} {\bibfnamefont
  {F.}~\bibnamefont {Jelezko}},\ }\href {\doibase
  10.1088/1367-2630/15/3/033027} {\bibfield  {journal} {\bibinfo  {journal}
  {New J. Phys.}\ }\textbf {\bibinfo {volume} {15}},\ \bibinfo {pages} {033027}
  (\bibinfo {year} {2013})}\BibitemShut {NoStop}%
\bibitem [{\citenamefont {Rose}\ \emph {et~al.}(2017)\citenamefont {Rose},
  \citenamefont {Tyryshkin}, \citenamefont {Riemann}, \citenamefont
  {Abrosimov}, \citenamefont {P.Becker}, \citenamefont {Pohl}, \citenamefont
  {Thewalt}, \citenamefont {Itoh},\ and\ \citenamefont {Lyon}}]{brendonArXiv}%
  \BibitemOpen
  \bibfield  {author} {\bibinfo {author} {\bibfnamefont {B.~C.}\ \bibnamefont
  {Rose}}, \bibinfo {author} {\bibfnamefont {A.~M.}\ \bibnamefont {Tyryshkin}},
  \bibinfo {author} {\bibfnamefont {H.}~\bibnamefont {Riemann}}, \bibinfo
  {author} {\bibfnamefont {N.~V.}\ \bibnamefont {Abrosimov}}, \bibinfo {author}
  {\bibnamefont {P.Becker}}, \bibinfo {author} {\bibfnamefont {H.-J.}\
  \bibnamefont {Pohl}}, \bibinfo {author} {\bibfnamefont {M.~L.~W.}\
  \bibnamefont {Thewalt}}, \bibinfo {author} {\bibfnamefont {K.~M.}\
  \bibnamefont {Itoh}}, \ and\ \bibinfo {author} {\bibfnamefont {S.~A.}\
  \bibnamefont {Lyon}},\ }\href {http://arxiv.org/pdf/1702.00504.pdf}
  {\bibfield  {journal} {\bibinfo  {journal} {arXiv:1702.00504 [quant-ph]}\ }
  (\bibinfo {year} {2017})}\BibitemShut {NoStop}%
\bibitem [{\citenamefont {Tyryshkin}\ \emph {et~al.}(2012)\citenamefont
  {Tyryshkin}, \citenamefont {Tojo}, \citenamefont {Morton}, \citenamefont
  {Riemann}, \citenamefont {Abrosimov}, \citenamefont {Becker}, \citenamefont
  {Pohl}, \citenamefont {Schenkel}, \citenamefont {Thewalt}, \citenamefont
  {Itoh},\ and\ \citenamefont {Lyon}}]{tyryshkin_electron_2012}%
  \BibitemOpen
  \bibfield  {author} {\bibinfo {author} {\bibfnamefont {A.~M.}\ \bibnamefont
  {Tyryshkin}}, \bibinfo {author} {\bibfnamefont {S.}~\bibnamefont {Tojo}},
  \bibinfo {author} {\bibfnamefont {J.~J.~L.}\ \bibnamefont {Morton}}, \bibinfo
  {author} {\bibfnamefont {H.}~\bibnamefont {Riemann}}, \bibinfo {author}
  {\bibfnamefont {N.~V.}\ \bibnamefont {Abrosimov}}, \bibinfo {author}
  {\bibfnamefont {P.}~\bibnamefont {Becker}}, \bibinfo {author} {\bibfnamefont
  {H.-J.}\ \bibnamefont {Pohl}}, \bibinfo {author} {\bibfnamefont
  {T.}~\bibnamefont {Schenkel}}, \bibinfo {author} {\bibfnamefont {M.~L.~W.}\
  \bibnamefont {Thewalt}}, \bibinfo {author} {\bibfnamefont {K.~M.}\
  \bibnamefont {Itoh}}, \ and\ \bibinfo {author} {\bibfnamefont {S.~A.}\
  \bibnamefont {Lyon}},\ }\href {\doibase 10.1038/nmat3182} {\bibfield
  {journal} {\bibinfo  {journal} {Nat. Mater.}\ }\textbf {\bibinfo {volume}
  {11}},\ \bibinfo {pages} {143} (\bibinfo {year} {2012})}\BibitemShut
  {NoStop}%
\bibitem [{\citenamefont {Julsgaard}\ and\ \citenamefont
  {M{\o}lmer}(2013)}]{julsgaard_fundamental_2013}%
  \BibitemOpen
  \bibfield  {author} {\bibinfo {author} {\bibfnamefont {B.}~\bibnamefont
  {Julsgaard}}\ and\ \bibinfo {author} {\bibfnamefont {K.}~\bibnamefont
  {M{\o}lmer}},\ }\href {\doibase 10.1103/PhysRevA.88.062324} {\bibfield
  {journal} {\bibinfo  {journal} {Phys. Rev. A}\ }\textbf {\bibinfo {volume}
  {88}},\ \bibinfo {pages} {062324} (\bibinfo {year} {2013})}\BibitemShut
  {NoStop}%
\end{thebibliography}
\end{document}